\documentstyle[11pt,aaspp4,flushrt]{article}

\def\msun{\ifmmode {\rm M_\odot} \else M$_\odot$\fi}
\def\msunyr{\ifmmode {\rm M_\odot~yr^{-1}}\else${\rm M_\odot~yr^{-1}}$\fi}
\def\lam{$\lambda$}
\def\lalpha{L$\alpha$}
\def\cm{\ifmmode {\rm cm} \else  cm \fi}
\def\cmmitwo{\ifmmode \rm cm^{-2} \else $\rm cm^{-2}$\fi}
\def\cmps{\ifmmode \rm cm~s^{-1}\else $\rm cm~s^{-1}$\fi}
\def\kmps{\ifmmode \rm km~s^{-1}\else $\rm km~s^{-1}$\fi}
\def\mdotav{\ifmmode \langle \dot m_o \rangle \else
$\langle \dot m_o \rangle$ \fi}
\def\eg{e.g.}

\def\cf{cf.}
\def\etal{et al.}

\begin{document}

\title{The Extraordinary Abundances of QSO Broad Absorption Line Regions:
A Matter of Novae?}

\author{Gregory A. Shields}
\affil{Department of Astronomy \\
The University of Texas at Austin \\
Austin, TX 78712}


\lefthead{Shields}
\righthead{QSO Broad Absorption Line Regions}

\begin{abstract}

The broad absorption lines (BALs) of QSOs indicate abundances
of heavy elements, relative to hydrogen, that are 1 to 2 orders of
magnitude higher than the solar values.  In at least one QSO,
an especially large enhancement of phosphorus is observed.  These abundances
resemble those in Galactic novae, and this suggests that novae may produce
the BAL gas. The needed rate of nova outbursts may come from single white
dwarfs that accrete gas as they pass through a supermassive accretion disk
around a central black hole.

\end{abstract}

\section{Introduction}

BAL QSOs  show broad absorption
troughs attributed to rapidly outflowing gas,
of unknown origin, located outside the continuum source and the broad
emission-line region. BALs typically occur in \lalpha,  C IV \lam 1549,
Si IV \lam 1400, N V \lam 1240, O VI \lam 1034, and sometimes
Mg II \lam 2798 and Al III \lam 1857 (see reviews by
Weymann, Turnshek, and Christiansen 1985, ``WTC''; and
Turnshek 1988, 1995).  BALs often set
in at the systemic velocity of
the QSO, but in some objects ``detached'' BALs begin at a
velocities up to $\sim10^4$ \kmps.  The absorption often
extends to velocities $\sim 3\times10^4$ \kmps,
with varying degrees of structure
and residual intensity through the line profile.  The derived column densities
for the absorbing ions  (e.g., C$^{+3}$) are $\sim10^{16}$~\cmmitwo.
BALs occur in about 10 percent  of radio quiet QSOs (Weymann
\etal~1991) and rarely in radio loud QSOs . Either most radio
quiet QSOs have
BAL material covering about 10 percent of the sky as seen from the continuum
source, or a subset of QSOs have BAL regions with a larger covering factor.
The BAL region may have a  disk-like geometry (Turnshek 1995;
Goodrich and Miller 1995).  The absorbing material
must have a transverse dimension
$\gtrsim 10^{16}$ cm, sufficient to cover the continuum source.  The N V BAL
often strongly absorbs the \lalpha\ emission line,  implying a radius
$\gtrsim 10^{18}$ cm (Turnshek \etal~1996).  The absorbing clouds, which
occupy only a tiny fraction of the volume,
may be accelerated by radiation pressure involving the
resonance lines or by ram pressure of a fast moving
wind (WTC). Proposed sources for the BAL gas include winds from red giant stars
(Scoville and Norman 1995) or from a supermassive
accretion disk (Murray \etal~1995).

\section{Chemical Abundances}
 
The BAL gas has remarkably high
abundances of heavy elements, relative to hydrogen.
The abundances are derived
from column densities of ions (in turn derived from
observed BAL optical depths)
together with photoionization models of the
ionization equilibrium.
Junkkarinen \etal\ (1987) argued for a minimum (Si/H)
$\geq$ 10(Si/H)$_\odot$ for their sample of BAL QSOs, and their discussion
suggests a likely value (Si/H) $\approx$ 25(Si/H)$_\odot$.   Turnshek \etal\
(1987) reported a C/H ratio 10 to 100 times the solar value in Q0932+501
(see also WTC); and for Q1413+113,
they found that S/C and either P/C or Fe/C are at least
100 times solar.  For Q0226-1024, Korista \etal\ (1992, 1995)
found a heavy element abundance
Z/Z$_\odot \approx$ 5 to 10, in the context of an assumed chemical
evolution scenario.  Analyzing the same
observations of Q0226-1024, Turnshek \etal\ (1996)
developed atwo component photoionization model that  gave ratios
$\sim$ 4, 3, 9, and 120 times solar for N/O, N/Si,
N/C, and N/H respectively (best fit model). Enhancements of other elements,
including S and Ar, are observed in some cases (Turnshek
1995).  Remarkably,
P/C is $\sim 65$ times the solar value in PG 0946+301 (Junkkarinen
\etal~1995; Junkkarinen 1995).  The heavy element
enhancements can be reduced, but not eliminated, by using complicated shapes
for the ionizing continuum (WTC).

Novae show very large enhancements
of C, N, O and sometimes Ne, Mg, Al,
Si, S, Ar, Ca, and Fe  (Andre\"a \etal~1994, and references therein).
Novae are produced by thermonuclear
explosions on the surface of white dwarf stars
accreting hydrogen rich gas from a close companion star
(see review by Starrfield 1989).
The ejected gas is enriched with heavy elements mixed in from the
white dwarf.
Enhancements of Ne and heavier elements,
observed in ``neon novae'', are attributed to
events on  O-Ne-Mg white dwarfs (Politano \etal~1995,
and references therein). Hydrogen
burning leads to especially large enhancements of nitrogen,
and in the case of O-Ne-Mg white dwarfs, heavier odd numbered elements.
Fig. 1  illustrates the abundances in a set of novae, determined
in a uniform manner by Andre\"a \etal\ (1994).
Abundances of C, O, and Si are typically enhanced by
one to two orders-of-magnitude, compared with solar values; and
N/H is enhanced by two to three orders-of-magnitude.

I have estimated abundances for several BAL QSOs by assuming that
their elemental abundance ratios  scale from those
given by Turnshek \etal\ (1996) for Q0226-1024 in
proportion to the
corresponding ionic column densities. (This assumes that
the ionization corrections are similar in all these objects.
Column density ratios for ions
of a single element are generally unavailable to constrain
models for individual objects.) Column densities were taken from
Junkkarinen \etal\ (1987) for 5 BAL QSOs and from
Junkkarinen (1995) and co-workers for Q0946+301.
Figure 1 shows the resulting
abundance ratios along with a value
(C/H) = 30(C/H)$_\odot$ for Q0932+501
(Turnshek \etal~1987).   The abundance
distributions for C, N, O, and Si for novae and BAL QSOs are similar.
Politano \etal\ (1995) predict
P/C ratios of $\sim$ 50 and 300 times solar for nova explosions on O-Ne-Mg
white dwarfs  with masses of
1.25 and 1.35 \msun, respectively, bracketing
the value for PG0946+301. These results suggest that the BAL gas may come
from novae in a massive star cluster in the active galactic nucleus.

High Si/C and P/C in BAL QSOs suggests that neon novae are
typically involved.  If O-Ne-Mg white dwarfs come only from progenitors of
mass 8---10 \msun\ (Nomoto 1984), they should be only a few
percent of Galactic white dwarfs.  Truran and Livio (1986)
attempted to explain the high observed incidence of neon novae in the Galaxy
in terms of a small accreted mass per outburst.  However, observed masses of
neon nova shells are $\sim 10^{-4}~\msun$ (e.g., Shore \etal~
1993), so another explanation may be needed.  If O-Ne-Mg white dwarfs come
from progenitors down to $\sim 5~\msun$ (Chiosi \etal~1989),
these would be a large fraction of all white dwarfs for a cluster age $\sim
10^8$ yr, the likely duration of a QSO episode (Norman and
Scoville 1988).  Alternatively, perhaps nuclear burning to
high atomic numbers, which Politano \etal\ (1995) find only
for the most massive white dwarfs, actually occurs for lower masses as well.
Finally, if the BAL geometry is such that debris of different novae are
comingled, as might occur in the disk geometry of Section 4, neon novae could
contribute  heavier elements to the mix.

The heavy element abundances in the broad emission-line gas of QSOs may be up
to an order of magnitude higher than solar, relative to hydrogen
(Hamann and Ferland~1993), but apparently are not so high
as in the BAL gas.
The emission-line abundances have been attributed to rapid  chemical evolution
involving the usual stellar sources of heavy elements (Hamann and Ferland
1993), and this would not explain high P/C.  This suggests that the two
regions likely have different origins.

\section{Ordinary Novae}

Novae are minor sources of interstellar gas in the Galaxy.  Can they
nevertheless be the dominant source of BAL gas?  We first consider
ordinary novae in a massive nuclear star cluster and then the
possibility of accretion onto single white dwarfs passing through
a supermassive accretion disk.

The mass of a nova shell typically is  $M_{ej} \approx
10^{-4.3}$~\msun\  (Warner 1989).  As the shell
expands and accelerates radially outward, it presumably fragments
into a collection of clouds or filaments.  We assume
that the lateral expansion continues at the original ejection velocity,
$v_{nov} \approx 2000~\kmps$, appropriate for neon novae (e.g.,
Ferland, Lambert, and Woodman 1977; Shore
\etal~1993; Gehrz
\etal~1985).  The column density of carbon atoms through the
debris is $N_C
\approx M_C/(\pi r_{sh}^2 m_c)$,  where $r_{sh}$ is the shell radius, $M_C$ is
the
mass  of carbon in the nova shell, and $m_C$ is the mass of a carbon atom.
Therefore, we may write
\begin{equation}
r_{sh} \approx (10^{17.3}~\cm) M_{C,-5}^{1/2} N_{C,16}^{-1/2},
\end{equation}
where $M_{C,-5} \equiv M_C/10^{-5}$~\msun\ and $N_{C,16} \equiv
N_C/10^{16}~\cmmitwo$.   For a carbon abundance $C/H = 10^{1.5} (C/H)_\odot$,
we expect  $M_C \approx 10^{-5.5}$ \msun.
Assuming that roughly half of the carbon is C$^{+3}$, we find
$N_{C,16} \approx 1$ for $r_{sh} \approx
10^{17.0} \cm$. This radius is large enough to cover the continuum source and,
within the uncertainties, the
\lalpha\ emitting region.  As the debris expand, the continuum
radiation pressure presumably ablates off material in the form of
small clouds, which move  radially outward as they are accelerated to the
observed outflow velocities (\cf\ WTC). For $\Delta R/r_{sh} \approx
w/v_{nov}$,
$N_{C,16}$  drops to
unity for distances $\Delta R \approx (w/v_{nov})r_{sh} \approx 10^{18}~\cm$
from an initial location at radius $R_{init}$, where $w \approx 10,000~\kmps$
is the outflow velocity.

The expected covering factor for nova shells depends on the nova rate
in the star
cluster in the galactic nucleus.  The covering factor for one shell is
$\Omega_{sh}/4\pi \approx \pi r_{sh}^2/(4\pi R^2) \approx 0.25
(v_{nov}/w)^2$,
where the second equality assumes $R_{init} < \Delta R$. The number of shells
in play at a given moment is
$N_{sh}
\approx \dot N R/w$, where  $\dot N$ is the rate of nova explosions and
$R/w \approx r_{sh}/v_{nov}$ is the crossing time.  We evaluate
$\Omega/4\pi$ for the value of $r_{sh}$ that gives $N_{C,16}\approx1$.
This leads to
\begin{equation}
{\Omega/ 4 \pi} \approx 0.25 \dot N r_{sh}
v_{nov} w^{-2}
\approx 10^{0.15} \dot N_o (v_{nov}/w) w_9^{-1}M_{C,-5}^{1/2}
N_{C,16}^{-1/2},
\end{equation}
where $\dot N_o$ is the nova rate per year and $w_9 \equiv w/(10^9~\cmps)$.
If $R_{init} \gtrsim \Delta R$, this result should be adjusted accordingly.
Note that $\Omega/ 4\pi$ varies as $(\dot N M_{ej}) M_{ej}^{-1/2}$, so that
events involving a small mass, such as novae, are highly effective
for a given average mass-loss rate, $\dot N M_{ej}$.  The nova rate in the
Galaxy (mass $\sim 10^{11}~\msun)$ is $\sim40~yr^{-1}$ (Warner
1989, Ciardullo
\etal~1990); and therefore we take $\dot N_o \approx
10^{-1.4} M_{cl,8}$, where
$M_{cl,8}$ is the mass of the nuclear star cluster in $10^8$~\msun.
Then we find
\begin{equation}
\Omega/4 \pi \approx 10^{-1.1} M_8  (v_{nov}/w) w_9^{-1}
\approx 10^{-1.8} M_8.
\end{equation}
A ten percent covering factor requires $M_8 \approx
10$.  Analogous estimates of the covering factor for BAL absorption due to
planetary nebulae, supernova remnants, and ``stellar contrails'' (Scoville and
Norman 1995) indicate that novae are competitive or
dominant by up to an
order of magnitude.  However, none of these sources match the BAL abundances
as naturally as do novae.

The mass loss rate in carbon alone can be estimated as
\begin{equation}
\dot M_C \approx  N_C m_C 4\pi R^2 (R/w)^{-1} (\Omega/4\pi)
\approx (10^{-4.4} M_\odot~yr^{-1}) N_{C,16} R_{18} w_9 (\Omega/4\pi).
\end{equation}
For $N_{C,16} \approx R_{18} \approx w_9 \approx 1$ and $\Omega/4\pi \approx
0.1$, this gives $\dot M_C \approx 10^{-5.4}$ \msunyr.  For $C/H
\approx 30(C/H)_\odot$, the total mass loss rate then is $\sim 10^{-4.2}$
\msunyr, a rather modest value.

Norman and Scoville (1988) discussed a coeval star cluster of mass
$\sim10^9~\msun$ whose evolutionary debris fuel the central
black hole. The age of the nuclear star
cluster is at most
$10^{9.3}$ yr at redshift $z=2$ for $q_o = 1/2$ and $H_o^{-1} = 15$ Gyr,
and it could be as young as the estimated QSO lifetime of $\sim10^8$ yr.
The ``evolutionary flux'' of
stars leaving the main sequence drops an order-of-magnitude as a
coeval cluster evolves from age $10^8$ to age $10^{9.5}$ years and somewhat
further by age $10^{10}$ years (Norman and Scoville 1988).
The effect of this on the nova rate is unclear.
The required star cluster, confined to a radius not
much larger than $10^{18}$ cm, would have a stellar velocity dispersion
of several thousand \kmps.  The stellar collision time  would
be shorter than the Hubble time and possibly even the estimated
$\sim 10^8$ yr lifetime of a QSO episode.  Observations
of nearby galactic nuclei offer little support for such massive nuclear star
clusters (\eg\ Lauer \etal~1992, and references therein).
Thus, we are  motivated to
consider ways to enhance the nova rate in QSOs.


\section{Single White Dwarfs}

An intriguing possibility
is suggested by the work of Artymowicz,
Lin, and Wampler (1993, ALW). They consider stars on orbits
passing through
an accretion disk around a central black hole of mass $M_h$, accreting disk gas
during each passage.
White dwarfs orbiting through the disk will also accrete disk material, and
this raises raises the possibility of nova explosions on single white dwarfs
that have accreted the requisite amount of hydrogen rich gas (\cf\ Truran
\etal~1977).  The circular velocity in units $10^8$~\kmps\
is $v_{c,8} \approx 10^{0.1} M_{tot,8}^{1/2} R_{18}^{-1/2}$, where $M_{tot}
= M_h + M_{cl}$.  The average stellar velocity, $v_*$, will be roughly $v_c$.
Let $v_{rel}$ be the velocity of a star relative to the orbiting material in
the disk.  If the cluster is
corotating with the disk, most stars will have $v_{rel}$
substantially less than $v_*$.  For parameters of interest, $v_{rel}$ is
less than the escape velocity from the surface
of a white dwarf and large compared with the sound speed in the
disk. The accretion rate onto the white dwarf while in the disk is
\begin{equation}
\dot m \approx 2.5\pi G^2 m^2 \rho v_{rel}^{-3}
\end{equation}
(Bondi and Hoyle 1944),
where $m \approx 1~M_{\odot}$ is the mass of the white dwarf
and $\rho$ is the gas density in the disk.
Following ALW, we assume that the disk thickness $H$
is such that the Toomre (1964) stability parameter, $Q$, is
near unity. Consequently,
$\pi \Sigma R^2/M_H \approx H/R \approx 10^{-2}$,
where $\Sigma = 2H\rho$ is
the disk surface mass density and a value $H/R \approx 10^{-2}$ corresponds to
the disk's expected vertical equilibrium.  Then
$\Sigma \approx (10^{2.8}~{\rm g}~\cmmitwo) M_{h,18} R_{18}^{-2}$,
$\rho \approx (10^{-13.5}~{\rm g~cm^{-3}}) M_{h,8} R_{18}^{-3}$,
and $\dot m_o \approx 10^{-10.1} M_{h,8} R_{18}^{-3} v_{rel,8}^{-3}$, where
$\dot m_o \equiv \dot m/1~\msunyr$.
If the star's
vertical velocity $v_z$ is  $\sim  v_{rel}/\sqrt{3}$, then the duration
of passage through the disk is $\Delta t \approx (10^{0.7}~{\rm yr}) R_{18}
v_{rel,8}^{-1}$, and the mass accreted is $\Delta m \approx (10^{-9.4}~\msun)
M_{h,8} R_{18}^{-2} v_{rel,8}^{-4}$.  If this much mass is accreted twice per
orbital period,\ $P \approx 2\pi R/v_* = (10^{3.3}~{\rm yr}) R_{18}
v_{*,8}^{-1}$,  then the average accretion rate is $\mdotav \approx
(10^{-12.4}) M_{h,8} R_{18}^{-3} v_{*,8} v_{rel,8}^{-4}$.
A typical white dwarf will
undergo an explosion at intervals $t_{nov} \approx (10^{8.1}~{\rm yr})
M_{h,8}^{-1} R_{18}^3 v_{*,8}^{-1} v_{rel,8}^4 $.

Suppose there are $N_{WD} \approx 10^{7} M_{cl,8}$ white dwarfs in
the nuclear star cluster (Allen 1973).  (For a Salpeter [1955]
initial mass function ranging from 0.2 to 20 \msun\ and a main sequence turnoff
of 1.4~\msun,
corresponding to an age of $10^{9.3}$ yr, one has $N_{WD} \approx
10^{6.8} M_{cl,8}$.) Let the stars have a typical velocity $\sigma
\leq v_*$ in the frame corotating with the disk. Then the nova rate for most
white dwarfs, with $v_{rel} \approx \sigma$, will be
$\dot N_o \approx 10^{-1.1} M_{h,8} M_{cl,8} R_{18}^{-3} v_{*,8}
\sigma_8^{-4}$.  With $M_h \approx M_{cl}$ and $v_*^2 \approx
GM_{tot}/R$, this gives $\dot N_o \approx 10^{-1.9} M_{tot,8}^{1/2}
R_{18}^{-3/2} (v_*/\sigma)^4$. For
$R_{18} \approx M_8 \approx 1$ and
$\sigma/v_*  \approx 0.5$, we then have
\begin{equation}
\Omega/4\pi \approx 10^{-0.8} (v_{nov}/w)  w_9^{-1} \approx 10^{-1.5}.
\end{equation}
However, accretion is most rapid onto stars with small $v_{rel}$, and these
can dominate the nova rate in spite of their relatively small numbers,
$N(v_{rel})/N \approx (v_{rel}/\sigma)^3$.  If relative
velocities down to some minimum, $v_{min}$, contribute to the nova rate,
then a simple integration over $v_{rel}$ shows that the preceeding expressions
for $\dot N_o$ and
$\Omega/4\pi$ should be increased by a factor $\sim(\sigma/v_{min})$.  If
violent nova explosions occur for $\dot m < 10^{-7}~\msunyr$
(van~den~Heuvel \etal~1992), then we may take $v_{min} \approx
10^{6.6}~\cmps \approx
10^{-1}\sigma$.  This increases $\dot N_o$ and $\Omega/4\pi$ by an order
of magnitude, so that $\Omega/4\pi \approx 10^{-0.5}$. Within the
uncertainties, this is consistent with the observed 10 percent incidence of
BALs. For the assumed parameters, novae involving single white dwarfs
outnumber ordinary novae by a factor $\sim 40$.

Indications of a flattened BAL region
are consistent with novae from white dwarfs with small
$v_{rel}/v_*$, which would orbit close to the disk ($z/R \approx
v_{rel}/v_c$).  The actual nova event would occur at a random place in the
white dwarf's orbit. The association of BALs with radio quiet QSOs might
involve the absence of a suitable accretion disk in radio loud objects,
perhaps because the latter typically occur in elliptical galaxies.

\section{Discussion}

In summary, the high chemical abundances derived for the BAL region of
QSOs resemble the abundances of Galactic novae.
The mass of heavy elements
in a nova shell is consistent with constraints on the dimensions and location
of the BAL absorbing material.  A very massive nuclear star cluster must be
postulated in order for ordinary nova outbursts to explain the observed
incidence of BALs in QSO spectra.  Accretion of gas by white
dwarfs passing through a supermassive accretion disk can provide sufficient
nova outbursts with a more modest star cluster.

Observational consequences include a small BAL region that could show changes
over a few years.  Different velocity components in the BALs of a given
object could result from different novae and show different abundances.  Odd
numbered elements such as Al and P should be especially enhanced in abundance.

\acknowledgments

This work was inspired in part by the talks by P. Artymowicz and V.
Junkkarinen (who mentioned high P in nova models) at the June, 1995, meeting
of the American Astronomical Society.  I am grateful to F. Bash, R. Ciardullo,
D. Lin, K. Korista, D. Lambert, S. Starrfield,  D. Turnshek,
and  J. C. Wheeler for valuable discussions, and to V. J. and D.
T. for communicating results in advance of publication.  This work
was supported in part by STScI grant GO-5434.03-93A.

\eject

\clearpage
\begin{figure}
\plotone{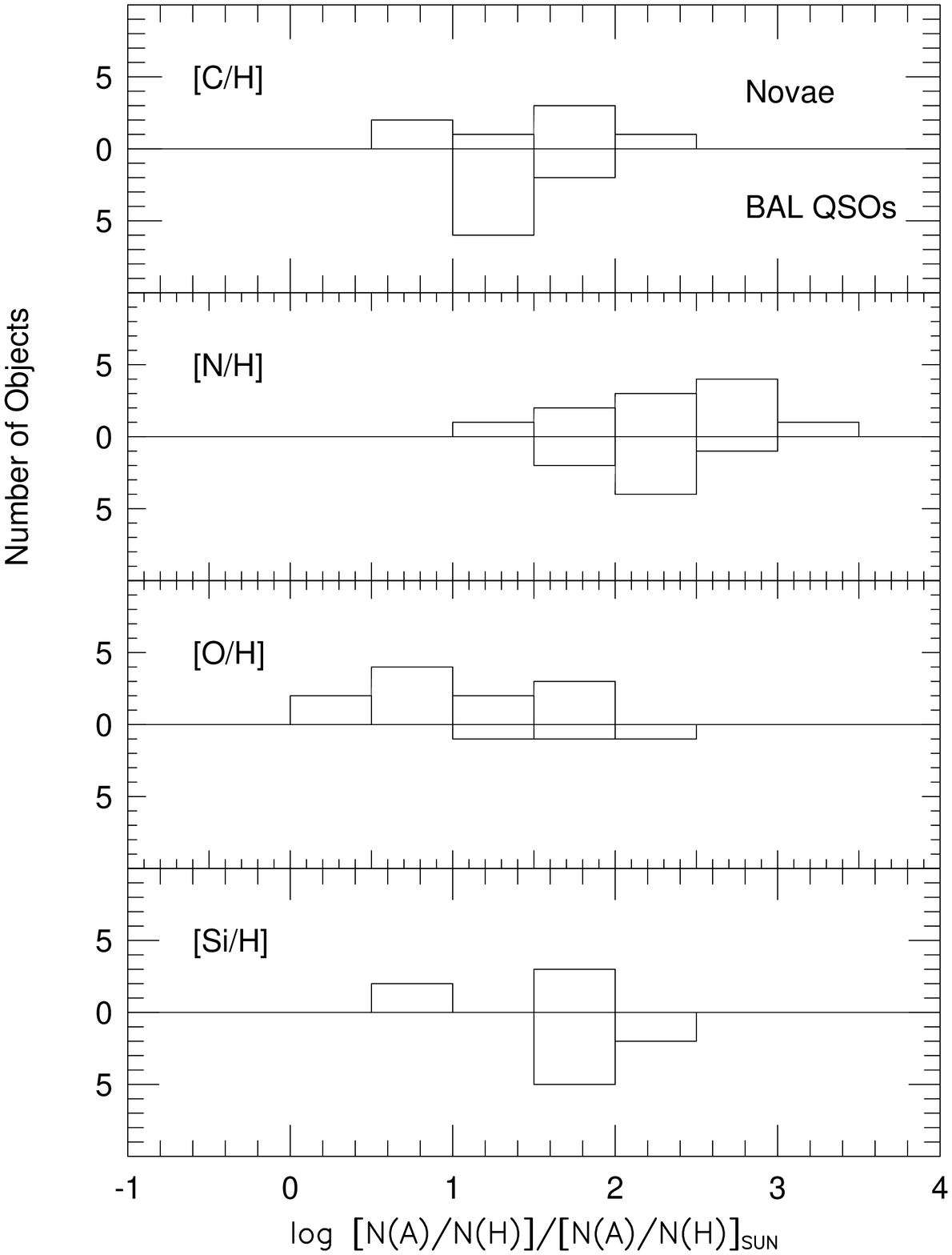}
\caption{
Histogram of C, N, O, and Si abundances relative to H
for novae and BAL QSOs, in bins of 0.5 dex, normalized to solar ratios.  In
each panel, novae are shown above the horizontal axis and BAL QSOs below it.
Each small tick on the vertical axis represents one object.
Results for novae are from Andre\"a \etal\ (1994), and those for BAL QSOs are
from several sources described in the text.
}
\end{figure}

\end{document}